\documentclass[aps,prl,twocolumn,amsmath,amssymb,floatfix]{revtex4}
\usepackage{graphicx,bm,color}
\newcommand{\eq}[1]{
\begin{equation}
{#1}
\end{equation}}

\newcommand{\E}{\bar{\varepsilon}}
\newcommand{\e}{\varepsilon}

\begin{document}
\title{Fermi-Edge Singularity in Chiral One-Dimensional Systems Far from Equilibrium}

\author{Iurii Chernii$^1$, Ivan P. Levkivskyi$^{2,3}$, Eugene V. Sukhorukov$^1$}
\affiliation{$^1$ D\'epartement de Physique Th\'eorique, Universit\'e de Gen\`eve,
CH-1211 Gen\`eve 4, Switzerland, \\
$^2$ Institute for Theoretical Physics, ETH Zurich, CH-8093 Zurich, Switzerland,\\
$^3$ Bogolyubov Institute for Theoretical Physics,
             14-b Metrolohichna Street, Kiev 03680, Ukraine
}

\begin{abstract}

We study the effects of strong coupling of a localized state charge to one-dimensional electronic channels out of equilibrium. While the state of this charge and the coupling strengths determine the scattering phase shifts in the channels, the nonequilibrium partitioning noise induces the tunneling transitions to the localized state. The strong coupling leads to a nonperturbative backaction effect which is manifested in the orthogonality catastrophe and the Fermi-edge singularity in the transition rates.
We predict an unusually pronounced manifestation of the non-Gaussian component of noise that breaks the charge symmetry, resulting in a nontrivial shape, and a shift of the position of the tunneling resonance.

\end{abstract}


%

\maketitle

\section{Introduction}

One-dimensional electronic systems are characterized by the increased role of interactions and the formation of gapless collective modes, that breaks down the Fermi liquid theory.
Instead, a state of matter is formed that is described as the Luttinger liquid \cite{LL}.
It has become experimentally accessible in the Quantum Hall (QH) effect systems, where the edge states act as one-dimensional (1D) chiral conductors.
Some of the recent exciting experiments worth mentioning would include:
electronic Mach-Zehnder (MZ) interferometry, in which the observed behavior of the interference oscillations has demonstrated the inapplicability of the single-electron description  \cite{heiblum interferometers 1};
energy relaxation along the QH edge channels with some unexplained losses in the energy transfer \cite{frederic pierre};
a demonstration of coherence and indistinguishability of independently emitted electrons \cite{feve};
and Coulomb interaction of a localized state, quantum dot (QD), with QH edge states \cite{experiment}, which presents a situation that potentially contains very interesting physics in the strong coupling regime. However, the latter experimental work \cite{experiment} has considered the system only in thermal equilibrium, where, due to the detailed balance, the whole range of implications of strong coupling has remained obscured.

\begin{figure}[!h]
\centering
\includegraphics[width=0.35\textwidth]{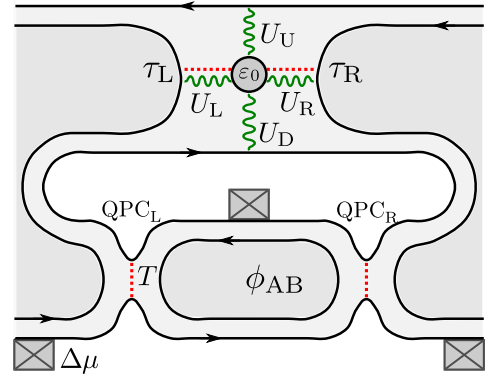}
\caption{Scheme of the experimental setup:
a Mach-Zender interferometer and a quantum dot are formed by the 1D electronic channels (solid lines with arrows) on the edges of the 2D electron gas (shaded areas) in the quantum Hall regime at filling factor $\nu=2$. The quantum dot with the single relevant level at $\e_0$ close to the Fermi energy is Coulomb interacting (waved lines) with the surrounding channels $a=U,D,L,R$, with coupling strengths $U_a$, and has a weak tunnel coupling with amplitudes $\tau_{\rm L,R}$ with the inner channels (dotted lines). Voltage bias $\Delta\mu$ is applied to one of the outer channels, which creates nonequilibrium excitations after the left partitioning quantum point contact (labeled $\textrm{QPC}_\textrm{L}$) with transparency $T$. The visibility of the interference pattern with respect to the Aharonov-Bohm phase $\phi_{\rm AB}$ is observed after the right QPC.
}
\label{fig:setup}
\end{figure}

In this work we extend the discussion to the case far from equilibrium, where the tunneling transitions recharging the QD are stimulated by the nonequilibrium processes in the edge channels.
The cooperation of the two factors, the nonequilibrium excitations in the channels and the backaction of the quantum fluctuations of the Fermi sea due to perturbation by an event of tunneling, is manifested in the strong coupling regime by the orthogonality catastrophe, leading to the Fermi edge singularity (FES) \cite{FES Mahan,FES Anderson,FES crowd} in the transition rates, and by their asymmetry with respect to the sign of the dissipative component of current. This is because the strong coupling negates the limitations of the {\it central limit theorem}, therefore giving access to the non-Gaussian fluctuations of the charge current that break the charge symmetry, while commonly they are small \cite{Sukhorukov-Edwards} and difficult to observe \cite{threshold-pekola,threshold-huarg}.

An equivalent system at zero temperature has been considered in the previous initiative by Rosenow and Gefen \cite{gefen-theory}. However, we must clearly indicate that in the specific regime defined above, our results are in sharp contradiction. Particularly, we believe that thoroughly accounting for the effects of strong coupling might have been an issue in their case, while they assumed that the coupling is finite (not weak) and is even maximum possible in the most interesting case.
In this respect, our results can also be contrasted with the results of some earlier works, in which the transition rates between a lead and a quantum dot interacting with noisy 1D conductors were studied, e.g., Refs.~\cite{Levinson,aleiner}.
In Levinson's work \cite{Levinson} the interaction between the QD and a QPC (analogous to the 1D edge channels in our case) was assumed to be weak. This assumption narrows the effects to the Gaussian noise,
and allowed a perturbative expansion in the interaction strength to the lowest order.
In turn, Ref.~\cite{aleiner}, although using strong coupling, 
has adopted an assumption about the structure of the scattering matrix, that the mixing of the states arriving from different reservoirs is weak.
This led to a substantially reduced effect of nonequilibrium noise, which only appeared as a small correction to the power-law exponent in the transition rates.
The differences between the assumptions and the findings of these works and ours essentially highlight the exceptionality of the situation where strong coupling is combined with the non-Gaussian noise.

It should be also noted that a number of exact solutions of the FES problem exist in various systems \cite{AL,abanin}, but they all require the use of highly complex methods that may obscure the physics, such as in Ref.~\cite{abanin} in which the 
distinctive nonequilibrium phenomenon of charge symmetry breaking in the strong coupling limit has been overlooked.
In 1D systems, however, the complexity can be largely overcome by employing the recently developed nonequilibrium bosonization technique \cite{noneq boson}. This approach naturally accounts for the interactions and allows us to reduce the problem of finding electron correlation functions to the calculation of the full counting statistics (FCS) of a 1D current \cite{levitov}, which can then be calculated analytically for asymptotically low transmission or reflection, or numerically in a general case.

\section{ The experimental setup}
The experimental setup is constructed by embedding the 1D channel as an arm of an electronic MZ 
interferometer at $\nu=2$ \cite{experiment,heiblum interferometers 1}, as shown schematically in Fig.~\ref{fig:setup}. Then the electron correlation function in the channel can be investigated by measuring the phase shift $\Delta \phi_{\rm AB}$ of the Aharonov-Bohm (AB) oscillations of the current $I = I_0  + I_{\Phi} \cos(\phi_{\rm AB})$ and  the visibility $V_0= I_\Phi/I_0$. The quantum dot charge fluctuates due to tunneling to some of the additional channels nearby. Voltage bias $\Delta\mu$ applied before the partitioning quantum point contact (QPC) with transparency $T$ creates nonequilibrium excitations in the arms of the interferometer.

The operating regime of the setup is determined by the interplay of the principal energy scales: the temperature $ \beta^{-1}$ of the environment, the quantum level broadening widths $|\tau_{\rm L,R}|^2\nu_F$ due to the tunnel coupling, and the classical level broadening $T \Delta\mu$ due to the nonequilibrium noise-induced transitions, where $\nu_F$ is the Fermi density of states and $\tau_{\rm L,R}$ are the tunneling amplitudes.
In the essentially quantum case, when $|\tau_{\rm L,R}|^2\nu_F$ is the dominant energy scale, the interferometer is found close to its ground state, where the interference pattern does not sustain any loss of visibility.

Assuming the low-energy limit, where the relevant energies: $\beta^{-1}$, the voltage bias $\Delta\mu$, and the detuning $\e_0$ of the quantum dot level from the Fermi energy are much smaller than the inverse time of flight through the interferometer, the internal dephasing in the channels can be neglected, and therefore the loss of coherence is solely due to the interaction with the localized state.
When the latter is occupied, the interference pattern acquires an additional shift compared to the empty state. This shift corresponds to the phase of scattering $2\pi\eta_{\rm D}$ on the localized charge.
Due to the transitions, this pattern is averaged with the occupation probabilities $P$, $1-P$ of the quantum dot, which gives the visibility $V$ and the phase shift $\Delta \phi_{\rm AB}$ as
\eq{
 V  e^{i \Delta \phi_{\rm AB}} = V_0 [ (1-P) + P e^{i 2\pi \eta_{\rm D}} ]
\label{visibility def}
,
}
where the constant prefactor $V_0$ is arbitrary, and without loss of generality we will further refer to the normalized value of the visibility $V$ in the sense of assuming $V_0=1$.

The experiment by Weisz {\it at al.}~\cite{experiment} has been realized in thermal equilibrium with $\beta^{-1}$ being the largest energy scale. In that case the transition rates at the impurity satisfy the detailed balance equation, and therefore the dot occupation probability $P=1/(1+e^{\beta\e_0})$  is Boltzmannian. This yields the result for the visibility $V  e^{i \Delta \phi_{\rm AB}} = (e^{i2\pi \eta_{\rm D}} + e^{\beta \e_0})/(1+e^{\beta \e_0})$ \cite{gefen-theory}. In the limit of strong Coulomb interaction and symmetric coupling $\eta_{\rm D} = \eta_{\rm U} = 0.5$ (see Fig.~\ref{fig:setup}), complete loss of visibility is observed, respectively, accompanied by the $\pi$-valued jump of the phase shift of the AB oscillations when the energy level of the dot crosses the Fermi level.

\section{ The model}
Throughout the rest of this paper we assume the low-temperature limit $\beta\Delta\mu\gg 1$, where the charge fluctuations at the impurity are activated by the partitioning noise of the $\rm QPC_L$.
The low-energy physics of a QH system shown in Fig.~\ref{fig:setup} can be conveniently described by the bosonized Hamiltonian (see, {\it e.g.}, Refs.~\cite{noneq boson,levkivskyi relaxation})
\begin{multline}
\label{hamiltonian}
\mathcal{H} = \int \frac{dxdy}{8\pi^2} \sum_{a b} \partial_x \phi_a(x) V_{a b}(x,y) \partial_y \phi_b(y) \\
+ \E_0 d^\dag d + d^\dag d  \int \frac{dx}{2\pi}  \sum_{a} U_a(x) \partial_x \phi_a(x)  \\
+ d^\dag \sum_a  \tau_a e^{i\phi_a(0)}   + {\textrm H.}{\textrm c.},
\end{multline}
where the Coulomb interaction potential $V_{a b}(x,y)$ of the charge densities $\rho(x) = \partial_x \phi_a(x)/2\pi$ at different points of channels $a$ and $b$ governs the propagation of the excitations along the edge channels; the bosonic fields $\phi_a$ satisfy the commutation relations $[\partial_x \phi_a(x) , \phi_b(y)] = 2i\pi \delta_{a b}\delta(x-y)$, and indexes $a,b$ enumerating the fields take values $\rm L,R,U,D$ on the corresponding channel to the left of, to the right of, up from, or down from the quantum dot. The bare level energy of the quantum dot $\E_0$ is determined by the applied gate voltage, which is the key controllable parameter of the system.
The quantum dot charge $d^\dag d$ interacts with the charge densities on the channels via the Coulomb potentials $U_a(x)$.
The amplitudes of tunneling from channel $a=L,R$ are denoted by $\tau_a$, and $e^{i\phi_a(0)}$ are the annihilation operators of an electron at $x=0$, chosen at the location of the quantum dot.

When the quantum level broadening is smaller than the classical broadening $|\tau_a|^2\nu_F \ll T\Delta\mu$, tunneling between the channels and the dot can be taken into account perturbatively.
The remaining part of the Hamiltonian can be diagonalized by eliminating the Coulomb interaction term with the help of a standard unitary transformation
$\tilde{\mathcal{H}} = e^{iS} \mathcal{H} e^{-iS}$, where
\eq{
S = d^\dag d \int dx \sum_a \sigma_a(x) \phi_a(x)
.
}
The functions $\sigma_a(x)$ are chosen with the aim to cancel the interaction part $ d^\dag d  \int dx  \sum_{a} U_a(x) \partial_x \phi_a(x)/2\pi $ of the Hamiltonian by the term produced after the transformation of the first line in Eq.~(\ref{hamiltonian}). This requirement is expressed by the integral equation
\eq{
U_a(x) = - \int dx' \sum_b V_{a b}(x,x')\sigma_b(x')
.
\label{integral equation}
}
On the other hand, the Coulomb potential along channel $a$, created by the impurity charge and the arbitrary charge density distribution $\rho_b(x)$ on channel $b$, is
\eq{
\varphi_a(x) =  U_a(x) d^\dag d  + \int dx' \sum_b V_{a b}(x,x')\rho_b(x')
,
}
from which it follows that the solution $\sigma_b(x)$ of Eq.\ (\ref{integral equation}) is nothing but the charge density $\rho_b(x) = \sigma_b (x)$, accumulated on the grounded channels $\varphi_a(x) = 0$, screening the charge present on the quantum dot, i.e., when $d^\dag d=1$. These densities are, naturally, localized in the interaction region around the quantum dot at $x=0$.
The total charges $\eta_a \equiv -\int dx \sigma_a(x) = \sum_b V_{a b}^{-1} U_b|_{k=0}$ can be expressed in terms of the zero-frequency Fourier components of the potentials $U_a(x)$.
In absence of other metallic objects in proximity of the dot, the electroneutrality principle implies that $\sum_a \eta_a = 1$. This relation repeats the statement of the Friedel sum rule \cite{friedel} via the direct connection between the additional phase $\Delta\phi_a = 2\pi \eta_a$ acquired by an electron passing by the dot and the charges $\eta_a$.

As a result of the transformation, the tunneling part of the Hamiltonian is rendered in the form
\eq{
H_t = \sum_b \tau_b d^\dag \ e^{i\phi_b(0) - i \sum_a \eta_a \phi_a(0)} + {\textrm H.}{\textrm c.},
\label{local tunneling}
}
where the fields $\phi_a(x)$ have been approximated by their value $\phi_a(x) \approx \phi_a(0)$ at $x=0$ since in the low-energy limit the fields change on distances longer than the size of the interferometer and thus are almost constant in the region of interaction.
The transformation also shifts the parameter $\E_0$, which determines the time evolution of the annihilation operator $d(t)$ by the static self-interaction energy of the accumulated charge density  $\e_0 = \E_0 + \sum_a \int dx U_a(x) \sigma_a(x)$ . The physical quantities, such as the visibility, will be investigated as functions of this parameter.

\begin{figure*}
\centering
\includegraphics[width=0.9\textwidth]{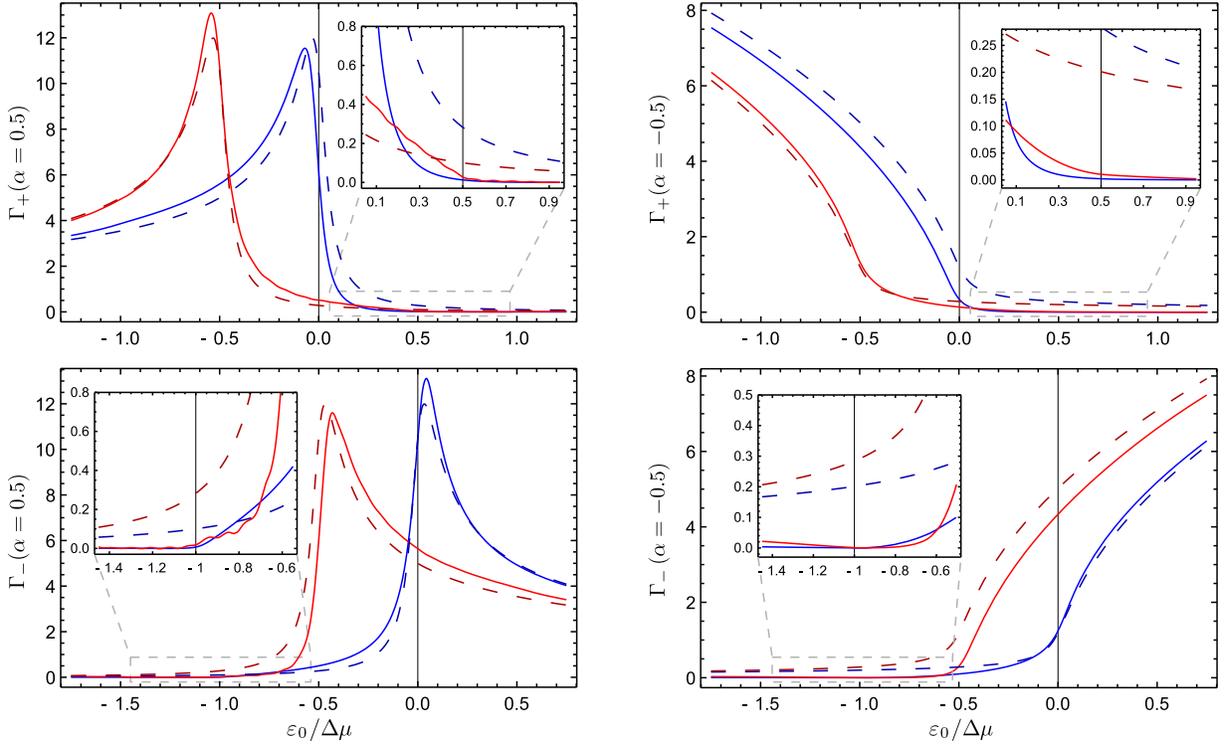}
\caption{(Color online) The transition rates $\Gamma_+$ and $\Gamma_-$ (in arbitrary units) as functions of the level energy parameter $\e_0$ for symmetrical screening $\eta_{\rm D}=0.5$ and exponent $\alpha=0.5$ (left) and $\alpha=-0.5$ (right) at different transparencies $T = 0.15$ (blue) and $T = 0.85$ (red). Analytic results in the Markovian limit are shown by the dashed lines, and the numerical data are shown by the solid lines.
As expected for screening $\eta_{\rm D}=0.5$, the positions of the corresponding peaks for $T\rightarrow 0 $ and $T\rightarrow 1$ are located at $\e_0=0$ and $\e_0=-\Delta\mu/2$, respectively.
While the analytic curves maintain consistent power-law tails on both sides, the nonequilibrium effects are represented by the characteristic suppression of the tails of the numerical curves; see, for example, $T=0.85$ at $\e = \Delta\mu/2$ (red line, top left panel and inset) and at $\e = - \Delta\mu$ for $T=0.15$ (blue line, bottom left panel and inset).
Note the symmetry (\ref{symmetry}) between the rates $\Gamma_+$ and $\Gamma_-$ for small and large transparency.
}
\label{fig:rates}
\end{figure*}

The rates of the tunneling transitions (Fig.~\ref{fig:rates}) between the quantum dot and the side channels are found from the \emph{Golden Rule} expression
\eq{
\Gamma_{\pm} =  \int\! dt \ \left\langle \mp \left|  H_t(0) H_t(t)  \right| \mp \right\rangle
\label{golden rule rates}
,
}
where $|-\rangle$ and $|+\rangle$ denote the states of the whole system unperturbed by the tunneling, when the quantum dot is empty and occupied, respectively.
Following the nonequilibrium bosonization approach \cite{noneq boson}, the fields $\phi_a(x_{\rm L},t)=-2\pi Q_a(t)$  are expressed in terms of the total charges $Q_a(t)$ that are transmitted across a certain cross section, such as right after the partitioning QPC for $\phi_{\rm D}(x_{\rm L},t)$ on the biased channel, or the corresponding cross sections of the other channels that are in the ground state.
Tunneling in the beam splitter QPCs is therefore taken into account nonperturbatively through the boundary conditions at $x=x_L$ (see also Ref.~\cite{safi}), and the correlation functions in Eq.~(\ref{golden rule rates}) can be expressed in terms of the correlation functions of these charges $Q(t)$ and thus via the FCS generating function \cite{levitov}
\eq{
\chi(\lambda, t) = \langle e^{i\lambda Q(t)} e^{-i\lambda Q(0)} \rangle
\label{FCS}
,
}
which is defined for $t>0$ and is continued for negative times as $\chi(\lambda,-t) = \chi^*(\lambda,t)$.
The total rates are therefore $\Gamma_\pm=\sum_b \Gamma_{b\pm}$, where $b=L,R$ and
\eq{\Gamma_{ b \pm}
 = |\tau_b|^2 \!\int\!\! dt \ e^{\mp i\e_0 t } \chi_b(\pm 2\pi(1-\eta_b),t) \prod_{a\neq b} \chi_a(\mp 2\pi \eta_a,t)
\label{rates through FCS}
.
}
Here $\chi_{a}(\lambda) \propto (i t + 0)^{-(\frac{\lambda}{2\pi})^2}$ for $a\neq D$ are the ground-state correlation functions on the unbiased channels. The dimensional prefactor is non-universal and will be omitted from here on.

For the nonequilibrium excitations from the partitioning QPC, the generating function is not trivial but can be found analytically in the long-time Markovian limit as a classical probability game result known as the Levitov-Lee-Lesovik formula \cite{levitov}:
\eq{
\log \chi_D(\lambda,t) = -\frac{\lambda^2}{4\pi^2} \log(i t + 0) + \frac{\Delta \mu t}{2\pi} \log \left(R+T e^{i\lambda}\right)
\label{levitov formula}
.
}
As a long-time asymptotic, it is applicable when the main contribution to the integrals in Eqs.\ (\ref{rates through FCS}) comes from the long times $t\sim 1/|\e_0 - \e^\star_0| \gg \Delta\mu^{-1}$, which requires small transparency $T\ll1$ or small reflection $1-T\ll1$, and small transmitted energies $|\e_0 - \e^\star_0| \ll \Delta\mu$. Here $\e^\star_0$ is the charge degeneracy point, i.e., the resonance energy, and will be discussed later.
One therefore obtains the result
\eq{
\label{rates analytic}
\Gamma_{b\pm}(\e_0)   \propto {\rm sgn}(\alpha_b) \textrm{Im}\left[ \frac{A_\pm}{(\varepsilon_0 + i\gamma)^{\alpha_b}} \right]
,}
where the opposite rates $\Gamma_{b+}$ and $\Gamma_{b-}$ are distinguished by the factors $A_+ = -1$ and $A_- = e^{i\pi\alpha_b}$, respectively.
The power-law exponent in the denominator is the conventional FES absorption-rate exponent
\eq{
\alpha_b = 1 -\Big[(1-\eta_b)^2 + \sum_{a\neq b}\eta_a^2\Big] = 2\eta_b - \sum_a \eta_a^2
,
\label{exponent}
}
expressed in terms of the accumulated charges, related to the scattering phases $2\pi\eta_a$ on the impurity.

%
%

The singularity is smeared by the real part 
of 
\eq{
\gamma = -\frac{\Delta\mu}{2\pi} \log\left[1 + T (e^{i2\pi \eta_{\rm D}}-1)\right]
\label{gamma}
,
}
while the imaginary part of $\gamma$ 
defines the shift of the level energy $\e_0$ depending on the transparency $T$ (see Fig.~\ref{fig:energy}) and determines the positions of the peaks of $\Gamma_{b\pm}(T)$, which are also discussed later.
In the free-fermion case $\alpha_b=0$ without the Coulomb interaction, the rates are reduced to the corresponding step functions.

Interaction with the same channel $a$, to which the tunneling occurs, provides the positive term $2\eta_b$ to the exponent $\alpha_b$, the so-called {\it Mahan contribution} \cite{FES Mahan}, which favors low-energy transitions, leading to a more singular character of the transition rate profiles as in Fig.~\ref{fig:rates} for the positive exponent $\alpha_b=0.5$. Interaction with other channels, $a\neq b$. reduces the exponent $\alpha_b$ by $\sum_a \eta_a^2$, representing the {\it Anderson contribution} \cite{FES Anderson}, which favors higher-energy transitions; see Fig.~\ref{fig:rates} for the negative exponent $\alpha_b=-0.5$, which is approximately the value in the experiment \cite{experiment}. However, since $\alpha_b$ depends on the distribution of the couplings $U_a(x)$ between the channels surrounding the quantum dot via the screening charges $\eta_a$, by switching from weak tunneling to weak backscattering of the inner channels, one can shift the balance and achieve positive values $\alpha_b >0$, changing qualitatively the energy dependence of the transition rates. Without loss of generality we will omit index $b$ in $\Gamma_{\pm}$ and $\alpha$, as in the case where there is tunneling to only one side channel.

\section{ Exact numerical calculations}

To complement the limited range of applicability of the Markovian approximation, we use the numerical data obtained in Ref.~\cite{numerics}.
It is valid for intermediate transparencies $0<T<1$ and shorter times $t \sim 1/\Delta\mu $.

The approach is based on the calculation of the FCS generator of the transmitted charges at the left QPC
\cite{noneq boson, levkivskyi relaxation}, expressed in terms of a free-fermion determinant (see Ref.~\cite{levitov}):
\begin{multline}
\langle e^{i\lambda Q(t)}e^{-i\lambda Q(0)}\rangle\ =\ \\
\det\{1-f(\varepsilon)+\exp[i\lambda P(t)\otimes S(\varepsilon)]f(\varepsilon)\},
\end{multline}
where $f(\varepsilon)$ is the electron distribution function, $P(t)$ is the projector on the time interval $[0,t]$, and $S(\varepsilon)$ is the scattering matrix of the QPC. Such a determinant can then be evaluated numerically \cite{numerics}.
Although the data are currently available for one special value of a phase shift of $2\pi\eta_{\rm D} = \pi$, that happens to be the value relevant for the existing experiment \cite{experiment}.

Since the Markovian approximation is valid for asymptotically small transparency or reflection, a direct comparison with the numerical calculations at finite transparency can not be expected. Nevertheless, we still see a surprisingly adequate quantitative agreement for the transition rate profiles at transparencies $T=0.15$ and $T=0.85$ (see Fig.~\ref{fig:rates}). While the shape of the curves remains very similar, note the shift of the resonance position of the solid lines compared to the dashed lines, which is related to the imaginary part of Eq.~(\ref{gamma}).

In the Markovian limit all the rates have identical energy profiles $\Gamma_\pm (T)$ and $\Gamma_\pm (1-T)$ (see dashed lines in Fig.~\ref{fig:rates}), but in the numerics (solid lines) the opposite direction rates for the same transparency $\Gamma_\pm (T,\e_0)$ are clearly distinct, emphasizing the specific correspondence
\eq{
\Gamma_+(T,\e_0) = \Gamma_-(1-T, - \e_0 -\Delta\mu/2)
\label{symmetry}
,
}
which reflects the particle-hole symmetry, while the simple charge symmetry is broken  $\Gamma_+(\e_0) \nleftrightarrow \Gamma_-(-\e_0)$. 

\begin{figure}
\centering
\includegraphics[width=0.4\textwidth]{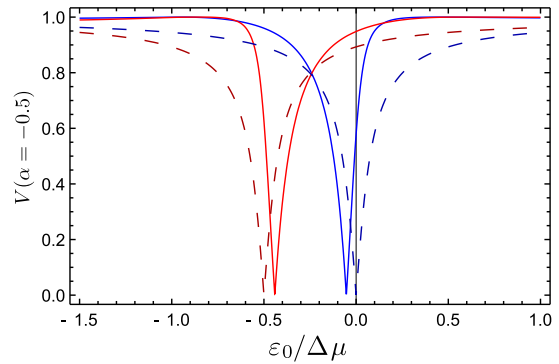}
\caption{(Color online) The visibility profiles found from the analytics (dashed lines) and numerics (solid lines) for transparencies $T=0.15$ (blue) and $T=0.85$ (red) for $\alpha=-0.5$ and screening $\eta_{\rm D}=0.5$ as in the experiment \cite{experiment}, corresponding to the phase shift of $\pi$ on the upper arm, hence the complete loss of visibility is achieved. Each analytic curve has a mirror symmetry and is also identical to its equivalent for $\alpha=0.5$ due to the symmetry incidental of the Markovian approximation. The numerical results manifest a well-pronounced asymmetry of the dip due to the non-Gaussian effects at strong coupling. }
\label{fig:visibility}
\end{figure}

The structure of the transition rates tails has an origin similar to that of the dynamic Coulomb blockade effect. In the present case it is sensitive to the sign of the charge of the edge excitations. Using the example of the $\Gamma_+$ rate at large partitioning transparency $1-T\ll1$ (solid red line in Fig.~\ref{fig:rates}), the holelike excitations on the arm of the interferometer attract the electron on the dot, assisting the energetically unfavorable tunneling transition from the side channel onto the dot and enhancing the tail at $\e_0>-\Delta\mu/2$, even compared to that of the Markovian limit result (dashed line). However, away from the resonance $|\e_0 - \e^\star_0| > \textrm{Re}(\gamma)$, the effects become perturbative in transparency $T$, and the tunneling is then enabled by single-particle excitations with a maximum energy of $\Delta\mu$; thus the tail is eventually suppressed at $\e_0=\Delta\mu/2$. Conversely, transitions onto the dot at low transparency $T\ll1$ are disadvantaged by the electronlike excitations, leading to the rapid decay of the $\Gamma_+$ rate at $\e_0>0$. The $\Gamma_-$ rates behave accordingly, in agreement with the symmetry~(\ref{symmetry}).

\begin{figure}
\centering
\includegraphics[width=0.45\textwidth]{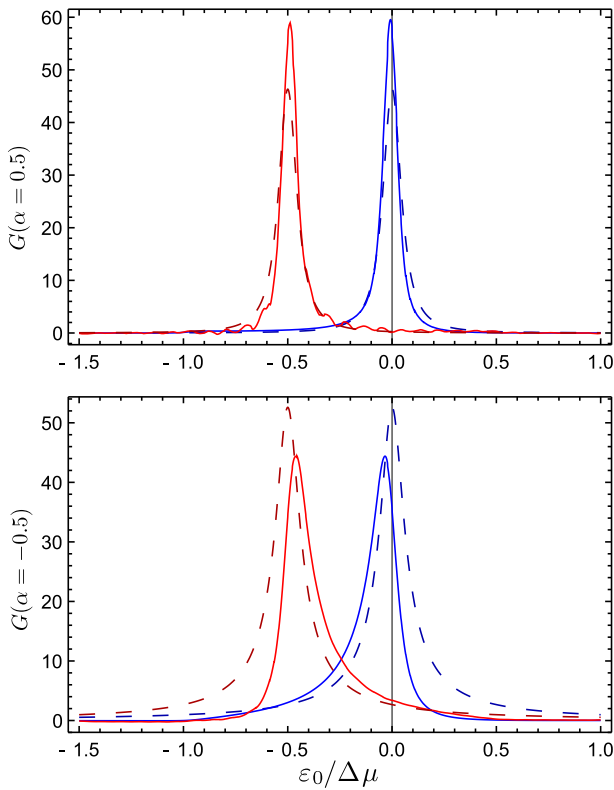}
\caption{(Color online) The tunneling conductance (in arbitrary) units for $T=0.15$ (blue) and $T=0.85$ (red) for the Markovian limit (dashed lines) and the numerics (solid lines) for (top) $\alpha = 0.5$ and (bottom) $\alpha = -0.5$.
Notice how the numeric curves hit the horizontal axis at $\e_0=\Delta\mu$ and $\e_0=-\Delta\mu/2$. For $\alpha=-0.5$ the Markovian limit breaks down considerably sooner, which explains the noticeable discrepancy between the numerics and analytic plots.
}
\label{fig:tunneling}
\end{figure}

\section{Visibility of the AB oscillations and tunneling currents}

The visibility is expressed (\ref{visibility def}) in terms of the stationary occupation probability $P=\Gamma_+/(\Gamma_+ + \Gamma_-)$, which is a ratio of the transition rates.
For $\eta_{\rm D}=1/2$ it demonstrates complete loss of coherence in the resonances (see Fig.~\ref{fig:visibility}) and is equivalent to the {\it detector function} discussed in Ref.~\cite{Sukhorukov-Edwards}.
The previous effect of the transition rates tails suppression is reincarnated in the rapid reconstruction of the visibility in the numerical results, shown in Fig.~\ref{fig:visibility}. We stress that unlike in Ref.~\cite{Sukhorukov-Edwards} in the single-photon, weak coupling regime, in our case the physical nature of this threshold effect is based on strong Coulomb coupling and is related to the finite energy $\Delta\mu$ carried by the partitioning excitations in the interferometer. The charge sign dependence of the coupling therefore leads to the pronounced asymmetry of the visibility dip seen in the numerical results due to the odd cumulants of the nonequilibrium noise.

In the long-time limit, however, the transition rates are insensitive to the charge sign and the branches of the visibility dip are symmetric (see dashed lines in Fig.~\ref{fig:visibility}).
For $\eta_{\rm D}=0.5$, the resulting visibility profile (\ref{visibility def}) becomes
\eq{
V e^{i\Delta\phi_{\rm AB}} =
1-\frac{2 \textrm{Im}\left[(\e_0 + i \gamma)^{-\alpha}\right]}{\textrm{Im}\left[\left(1+e^{i \pi (1-\alpha)}\right) (\e_0 + i \gamma)^{-\alpha}\right]}
\label{visibility mark}
.
}
The branches of the visibility dip exhibit a crossover between the two asymptotes:
$V\sim 2|\e_0-\e^\star_0|/\pi\theta_1$ with the effective temperature $\theta_1 =  \textrm{Re}\gamma (2/\pi \alpha) \tan( \pi\alpha/2)$ at $|\e_0-\e^\star_0| \ll T\Delta\mu $, and $V\sim 1-2\theta_2/\pi|\e_0-\e^\star_0|$ with $\theta_2 = \textrm{Re}\gamma [\pi\alpha/\sin(\pi\alpha)]$ at $ T\Delta\mu \ll |\e_0-\e^\star_0| \ll \Delta\mu$.
Both temperatures $\theta_1$ and $\theta_2$  become equal to the ``free-fermion'' noise temperature $\textrm{Re}\gamma$ of the weak coupling limit $|\alpha| \ll 1$ when the transition rates become regular Lorentzians \cite{Sukhorukov-Edwards}.
Note that even at fixed upper channel screening $\eta_{\rm D}$ and a phase shift of $\pi$, the exponent $\alpha$ can still vary due to the remaining couplings.

Since the power-law factor $(\e_0 + i \gamma)^{1-\alpha}$ enters homogeneously to the transition rates (\ref{rates analytic}), it is canceled in the tails of the visibility profile (\ref{visibility mark}).
Moreover, one can demonstrate that for the Markovian limit there is a symmetry $V_M(\alpha) = V_M(-\alpha)$, even though the transition rate profiles for $\alpha=0.5$ and for $\alpha=-0.5$ are unmistakably distinct, as can be seen in Fig.~\ref{fig:rates}.
This observation hints that the visibility itself is not the most fully characteristic quantity for an experimental consideration.
We therefore suggest that apart from the visibility, other measurements can be very interesting, such as the linear-response tunneling current (\ref{linear current}) through the quantum dot.
For this, a small bias $\delta \mu$ is applied to one of the tunneling contacts, and the linear response tunneling current $I_\beta= P_0\Gamma_{\beta+} - P_1\Gamma_{\beta-}$ through the quantum dot is measured
\eq{
\label{linear current}
I_{\rm L}=\frac{\Gamma_{\rm L+}\Gamma_{\rm R-} - \Gamma_{\rm R+} \Gamma_{\rm L-}}{\Gamma_{\rm R+}+ \Gamma_{\rm R-} + \Gamma_{\rm L+}+ \Gamma_{\rm L-}}
.
}

Assuming symmetric tunneling couplings $\tau_{\rm L}=\tau_{\rm R}=\tau$ and charge screening $\alpha_{\rm L} = \alpha_{\rm R} = \alpha$, the resulting differential conductance $G_{\rm L}=\partial I_{\rm L} / \partial \delta\mu$ then retains the signature of the FES power law in the Lorentzian-like peak
\eq{
G \propto
\frac{\varepsilon_F^{\alpha-1}}{\left| \e_0 + i \gamma \right| ^{\alpha+2}} \frac{ \textrm{Re}(\gamma) }{\cos \left\{ \alpha \left[\frac{\pi }{2}-\arg (\e_0 + i \gamma)\right]\right\} }
,
}
with the power-law tails $|\e_0-\e^\star_0|^{\alpha+2}$ and the resonance width $\textrm{Re}\gamma = -\Delta\mu/2\pi\log|R+T e^{i2\pi\eta_{\rm D}}|$ due to the nonequilibrium noise (see Fig.~\ref{fig:tunneling}). However, due to the suppression of the tails at $|\e_0-\e^\star_0|\sim\Delta\mu$ indicated by the numerical results, the power law spans only the limited segment of $T\Delta\mu < |\e_0-\e^\star_0| < \Delta\mu$, which can prove to be problematic for getting a good fit at moderately small transparencies $T$.

\begin{figure}
\centering
\includegraphics[width=0.4\textwidth]{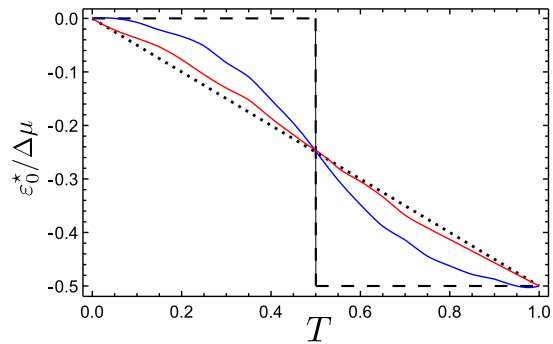}
\caption{(Color online) The position $\e^\star_0/\Delta\mu$ of the visibility dip center as a function of the QPC transparency. Solid lines are the numerical results for $\alpha=0.5$ (blue) and $\alpha=-0.5$ (red). The dashed line is for the Markovian limit, although it is applicable only asymptotically for $T\rightarrow 0$ and $T\rightarrow 1$. The dotted line presents the Gaussian case with suppression of the higher cumulants, reflecting simply the average charge density on the channel as the partitioning QPC transparency changes.
}
\label{fig:energy}
\end{figure}

Alternatively, spectroscopy of the quantum dot can be performed by filling or draining the dot with a finite bias in one lead at highly asymmetric tunneling couplings (see, for example, the recent experiment in \cite{spectroscopy}) and measuring directly the transition rates in one of the contacts. Then the exponent can be accessed from the nonvanishing tails, and the characteristic shape of the decaying tails should also be observable.

A particularly notable feature, related to the phase transition studied in Ref.~\cite{noneq boson}, is the dependence of the position $\e^\star_0$ of the resonance, where the visibility is maximally reduced, on the transparency $T$ of the partitioning QPC $T$.
At full transparency, the arm of the interferometer acts as an additional gate with the voltage $\Delta\mu$, which electrostatically raises the energy level of the quantum dot by $\eta_{\rm D} \Delta \mu$. The resonance therefore occurs at a lower value of the original parameter $\e^\star_0 = - \eta_{\rm D} \Delta \mu$. This corresponds to the shift obtained in the Markovian approximation as the imaginary part $\textrm{Im}\gamma$ of Eq.~(\ref{gamma}). For $\eta_{\rm D}=0.5$ the argument of the logarithm is real, and $\textrm{Im}\gamma$ jumps from zero for $T<0.5$ to $-\Delta\mu/2$ for $T>0.5$ \cite{ftnt arbitrary coupling}.
If the higher cumulants in the correlation function (\ref{FCS}) are suppressed for some reason, one can also consider the Gaussian approximation that predicts a linear drift of the level energy on the QPC transparency (dotted line in Fig.~\ref{fig:energy}).
The numerical results show a behavior that is somewhat intermediate between the Markovian and the Gaussian limits, suggesting that for $\alpha=0.5$ the higher cumulants have more influence than they do for $\alpha=-0.5$. It is therefore particularly interesting that the actual profiles are measured in an experiment.

\section{Summary}
A quantum Hall edge channel, embedded in an electronic interferometer, with a controllable Coulomb interaction with an artificial impurity, is a very promising system for investigating the interaction and noise effects on the Fermi-edge singularity manifestations in a one-dimensional electronic system.
In the low-energy limit, in which there is no intrinsic loss of coherence on the edge channels, the visibility of the interference pattern is only suppressed due to the averaging of the phase of scattering on the fluctuating impurity charge (\ref{visibility def}).
Experimental efforts to study such a system have been undertaken recently \cite{experiment}, although they concentrated on the thermal equilibrium case, in which the observed visibility is trivial due to the thermal occupation of the impurity.
We extend the discussion to the case where the transitions at the impurity are induced by the nonequilibrium partitioning noise created in the interferometer and the backaction of the Fermi sea perturbations due to tunneling. Both factors are strong and are taken into account nonperturbatively.
The nonequilibrium bosonization technique \cite{noneq boson} is the framework of choice, which allows us to express all the electron correlation functions in terms of the full counting statistics (\ref{FCS}) of the charge transmitted through the beam splitter. The analytical expression for the FCS (\ref{levitov formula}) is used in the Markovian limit for weak tunneling or backscattering in the partitioning QPC, and for intermediate transparencies the FCS is computed numerically.

We have provided a comprehensive description of the system in the given regime.
Even in the long-time, Markovian limit, the visibility profile (Fig.~\ref{fig:visibility}) is found to nontrivially depend on both the FES exponent (\ref{exponent}) and the parameters of the nonequilibrium noise, which are determined by the partitioning QPC transparency.
Beyond the long-time limit, as revealed by the numerical calculations, we discover a prominent manifestation of the non-Gaussian component of noise. In fact, at strong Coulomb coupling this becomes a dominant effect and leads to a particular kind of particle-hole symmetry, which also causes a pronounced asymmetry between the tunneling rates at small transmission versus small reflection of the QPC. Consequently, the non-Gaussian effects appear in the asymmetry of the visibility dip branches and the characteristic dependence of the dip position on the QPC transparency (Fig.~\ref{fig:energy}).
Considering all the predicted features, we can strongly recommend further experimental investigations with this type of setup operating out of equilibrium.

\begin{acknowledgments}
We thank M. Zvonarev for fruitful discussions, and we acknowledge the support from the Swiss National Science Foundation.
\end{acknowledgments}


\begin{thebibliography}{5}
\bibitem{safi}
A different approach was formulated earlier in Ref.~\cite{safi-ref}
in the context of the dynamical scattering theory for bosons, in which boundary fields are replaced by time-dependent electrochemical potentials with the purpose of finding the linear-response conductance matrix.
\bibitem{ftnt arbitrary coupling}
For Coulomb coupling other than $\eta_D = 1/2$, however, there will be a finite level energy shift for all transparencies, smearing the dotted line in Fig.~\ref{fig:energy}, and no complete loss of coherence.
\end{thebibliography}
\end{document}